\documentclass[a4paper]{jpconf}
\usepackage{graphicx}

\newcommand{\Saclay}{DAPNIA, Centre d'\'Etudes Nucl\'eaires de Saclay (CEA-Saclay), Gif-sur-Yvette, France}
\newcommand{\Saclayx}{$^{a}$}
\newcommand{\APC}{APC, Université Paris 7 Denis Diderot, Paris, France}
\newcommand{\APCx}{$^{b}$}
\newcommand{\LAL}{Laboratoire de l'Acc\'el\'erateur Lin\'eaire, Orsay, France}
\newcommand{\LALx}{$^{e}$}

\newcommand{\Ioannina}{University of Ioannina, Ioannina, Greece}
\newcommand{\Ioanninax}{$^{h}$}
\newcommand{\Dortmunt}{University of Dortmunt, Dortmunt, Germany}
\newcommand{\Dortmuntx}{$^{g}$}
\newcommand{\Thessaloniki}{Aristotle University of Thessaloniki, Greece}
\newcommand{\Thessalonikix}{$^{d}$}
\newcommand{\Athens}{National Center for Scientific Research ``Demokritos'', Athens, Greece}
\newcommand{\Athensx}{$^{c}$}
\newcommand{\Zaragoza}{Instituto de F\'{\i}sica Nuclear y Altas Energ\'{\i}as, Universidad de Zaragoza, Zaragoza, Spain }
\newcommand{\Zaragozax}{$^{f}$}

\newcommand{\Finland}{CUPP, Pyhasalmi, Finland}
\newcommand{\Finlandx}{$^{i}$}

\begin{document}
\title{Progress on a spherical TPC for low energy neutrino detection}

\author{S~Aune$^{a}$, P~Colas$^{a}$, H~Deschamps\Saclayx, J~Dolbeau\APCx, G~Fanourakis\Athensx,
E~Ferrer~Ribas\Saclayx, T~Enqvist\Finlandx, T~Geralis\Athensx,
Y~Giomataris\Saclayx, P~Gorodetzky\APCx,
G~J~Gounaris\Thessalonikix, M~Gros\Saclayx, I~G~Irastorza
\footnote{attending speaker: Igor.Irastorza@cern.ch} \Saclayx,
K~Kousouris\Athensx, V~Lepeltier\LALx, J~Morales\Zaragozax,
T~Patzak\APCx, E~A~Paschos\Dortmuntx, P~Salin\APCx,
I~Savvidis\Thessalonikix \ and J. D. Vergados\Ioanninax }

\address{\Saclayx\Saclay \\ \APCx\APC \\ \Athensx\Athens \\ \Thessalonikix\Thessaloniki \\
\LALx\LAL \\ \Zaragozax\Zaragoza \\ \Dortmuntx\Dortmunt \\
\Ioanninax\Ioannina \\ \Finlandx\Finland}

\begin{abstract}
The new concept of the spherical TPC aims at relatively large
target masses with low threshold and background, keeping an
extremely simple and robust operation. Such a device would open
the way to detect the neutrino-nucleus interaction, which,
although a standard process, remains undetected due to the low
energy of the neutrino-induced nuclear recoils. The progress in
the development of the fist 1 m$^3$ prototype at Saclay is
presented. Other physics goals of such a device could include
supernova detection, low energy neutrino oscillations and study of
non-standard properties of the neutrino, among others.
\end{abstract}

\section{The spherical TPC concept and first prototype}

The spherical TPC is a novel concept \cite{Giomataris:2003bp} with
very promising features, among which is the possibility of easily
instrumenting large target masses with very low energy threshold.
This could open the way of detecting the tiny (few hundreds of eV)
nuclear recoils produced by neutrino-nucleus coherent interaction,
which, although a standard process, has never been within reach of
current detector´s sensitivities.

The spherical TPC consists of 2 concentric spheres, the external
one usually connected at ground and the inner one at high
potential. The external sphere plays also the role of the vessel
that tightly encloses the target gas inside the drift volume. The
ionization charges produced in the interaction drift towards the
center are collected by an adequate gaseous readout, which covers
the surface of the inner sphere. In the simplest design, such
amplification structure is just a small spherical electrode,
around which the avalanche is produced, and which is read by a
single channel electronic chain. More sophisticated options are
envisaged for future prototype´s readouts, the preferred choice
being Micromesh Gaseous Structures (Micromegas
\cite{Giomataris:1995fq}) due to the high precision, fast response
and excellent energy resolution. High efficiency for detecting
single electrons have been proved with Micromegas
\cite{Derre:1999wh} even at high pressures \cite{Gorodetzky}. In
addition, Micromegas readout is currently being used for solar
axion detection in the CAST experiment
\cite{Andriamonje:2004hw,Andriamonje:2004hi} where a great
stability and ability to reject background events has been
achieved.

A first prototype has been built in Saclay to demonstrate the
concept. The external sphere is a 1.3 m diameter, 6 mm thick
copper vessel, and the central amplification electrode is a sphere
of 14 mm. First tests with Ar/CO$_2$ and Ar/Isobutane mixtures
have proven issues like robustness, gain and stability of
operation \cite{idm2004,lrt2004}.

In spite of the simplicity of the readout, some spatial resolution
is achieved in the radial coordinate by inspecting the time
pattern of the charge pulse detected at the center of the sphere.
Its temporal extension is determined by the longitudinal diffusion
of the ionization cloud and therefore by the distance drifted. The
$1/r^2$ dependence of the electric field enhances this effect with
respect to plane or cylindrical TPCs. Preliminarily, a resolution
of at least 10 cm has been achieved in the estimation of the
radial coordinate, and better values will be achieved when a
faster readout structure, like Micromegas, will be implemented.
This will also allow to have transversal dispersion information,
by means of an appropriate pixelization of the Micromegas readout.
Nevertheless, the result already achieved by the current prototype
is enough to perform rough fiducial cuts and to achieve some
degree of event identification and background rejection.

\begin{figure}[t] \center
  \includegraphics[height=.15\textheight]{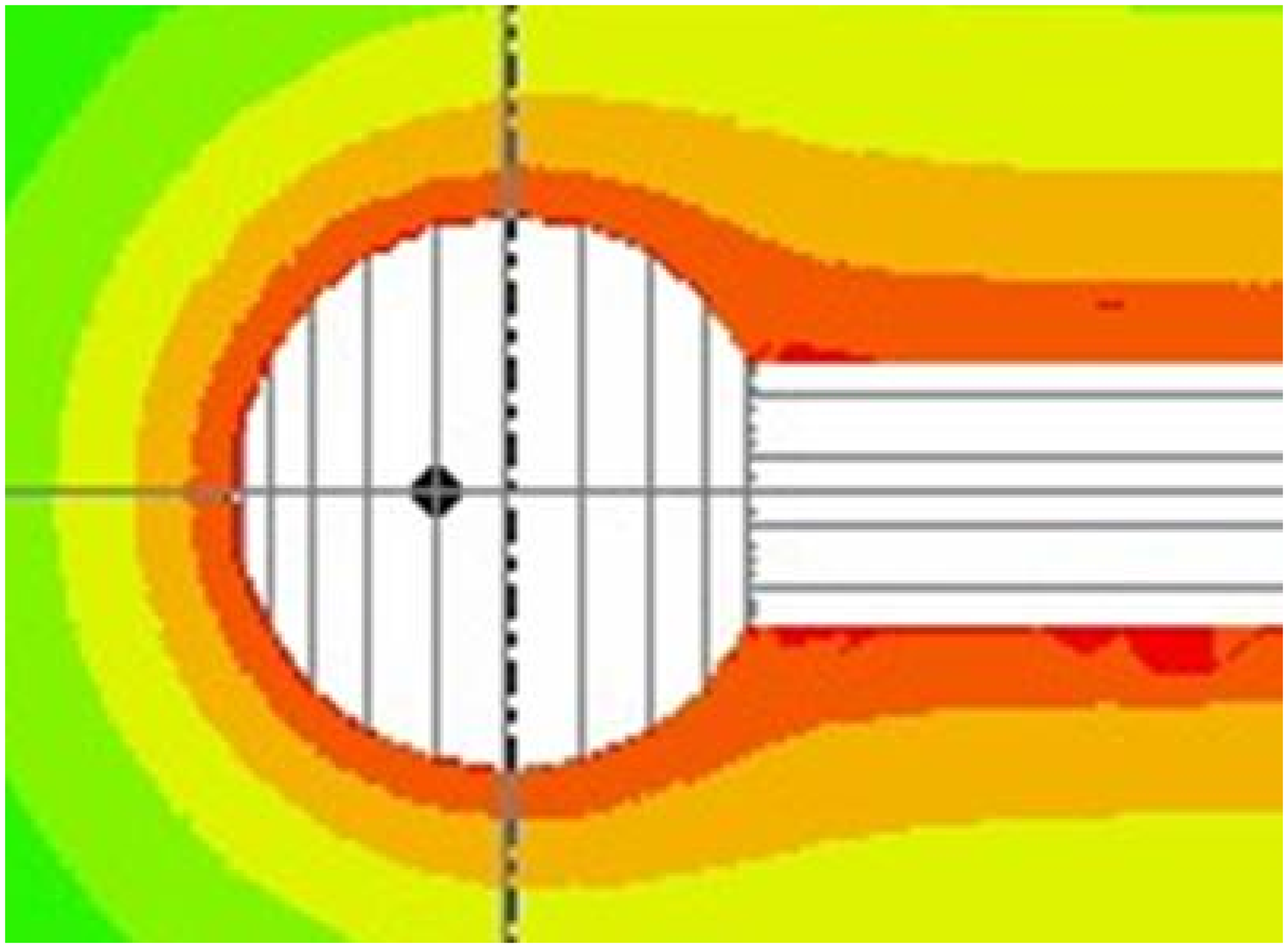}
  \includegraphics[height=.15\textheight]{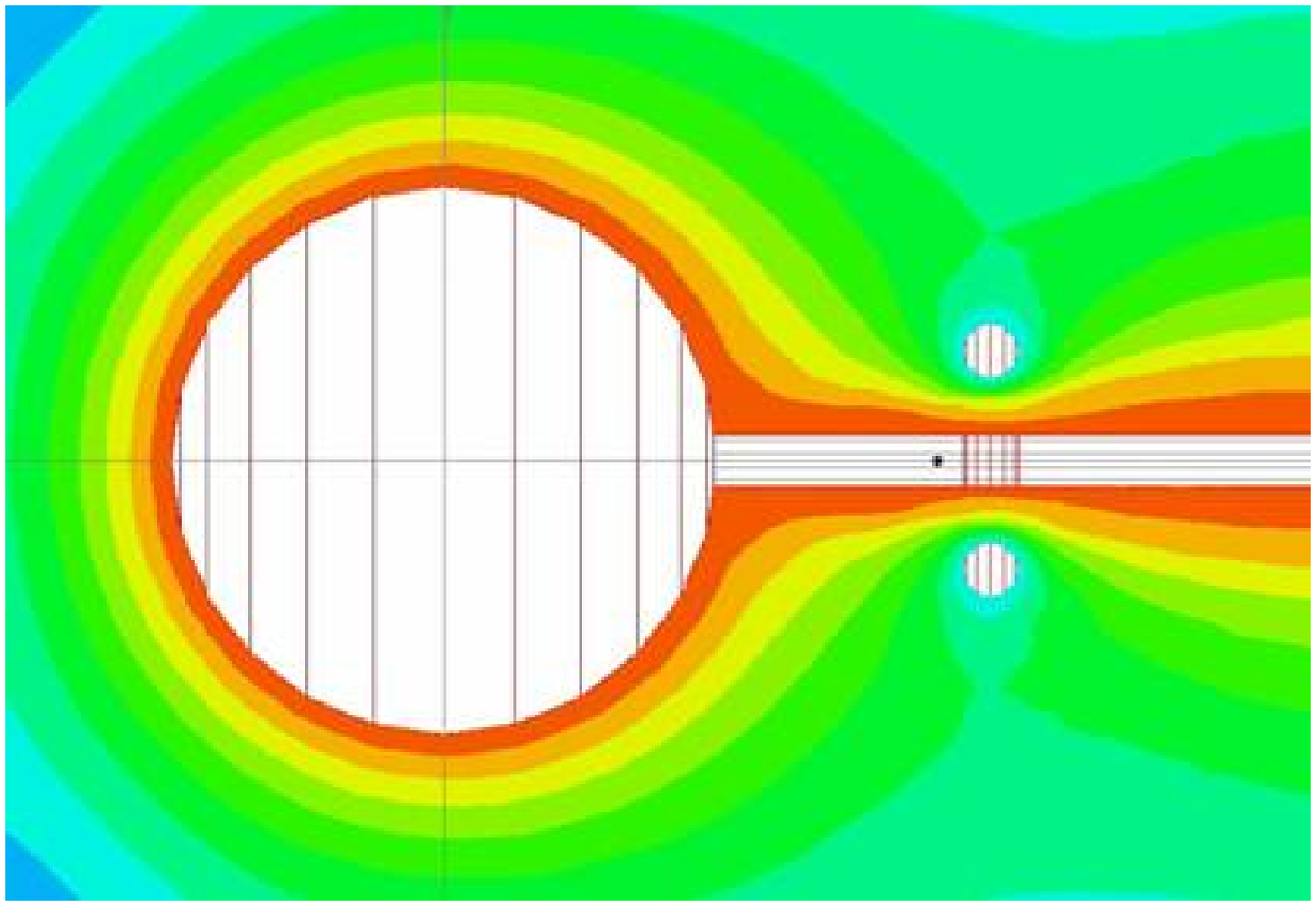}
    \includegraphics[height=.15\textheight]{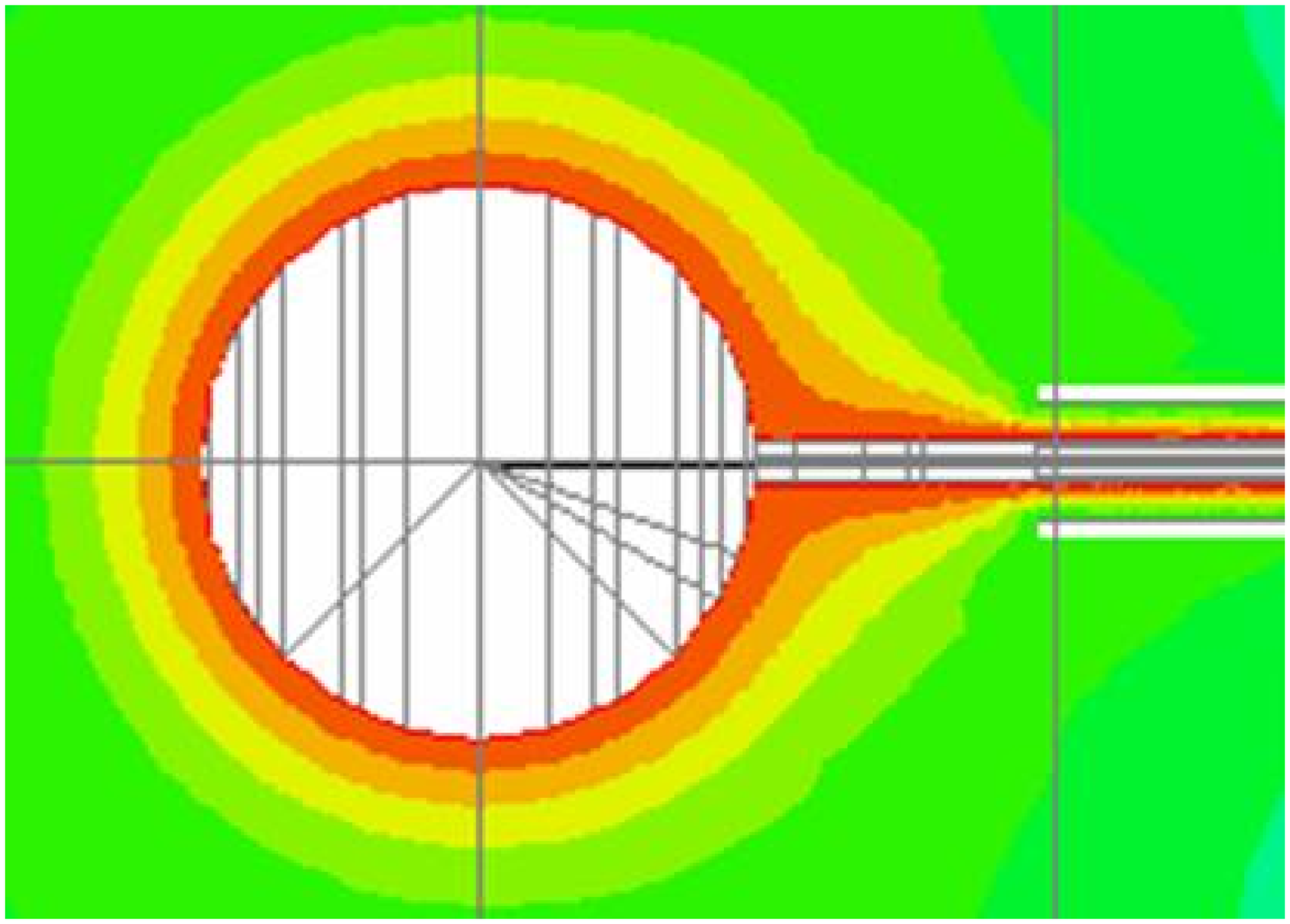}
     \caption{Numerical calculation of the electric field around the
 spherical electrode both for the single stick configuration (left)
 and for the case of a correcting ring (center) or cylinder (right).
 \label{field}}
\end{figure}

Other important advantages of the spherical TPC are:

\begin{itemize}
  \item The spherical geometry naturally focuses a large drift volume
  into a small amplifying
detector with only a few (or even just one) read-out channels. It
is the most cost-effective way of instrumenting a large detector
volume with a minimum of front-end electronics. Such approach
simplifies the construction and reduces the cost of the project.
  \item Spherical symmetry minimizes the external surfaces per unit of
  detector volume, as well as the thickness of material needed to hold
  the gas, therefore allowing a lower background per unit volume
  due to external surface or material contaminations.
  In principle, the simplicity of its design should
  allow an easy optimization from the point of view of
  radiopurity. Work is in progress to evaluate expected sources of
  backgrounds as well as to design a radiopure version of the
  present prototype.
    \item Large drift volumes can be
  built without the use of a field cage, unlike cylindrical TPCs.
The spherical symmetry means that the detector capacity is
  very low, allowing for extremely low levels of electronic noise
  (in fact, the outer sphere acts as a perfect Faraday cage to the inner electrode).
  The first preliminary tests with the prototype have easily
  achieved estimated thresholds of about 200 eV. Keeping in mind
  that no special measure has been taken for reducing electronic
  noise, using special quiet electronics or pushing to very high
  detector gains, the prospects to achieve thresholds well below
  100 eV are realistic.
\end{itemize}

Current efforts focus on the design of an electrostatic structure
that allows to bring the high voltage to the internal sphere with
minimal distortion of the spherical field, both for purposes of
drift and homogeneous amplification all around the small sphere.
In the absence of such structure, the central stick bringing the
high voltage to the small sphere (and mechanically supporting it)
distorts the field away from the ideal spherical field. Numerical
calculations, like the ones shown in Fig. \ref{field}, show that
approximately only one third of the detector volume has an
electric field reasonably close to the ideal one. To correct for
this effect several basic ideas are pursued, both by doing
numerical calculations of the electric field and by performing
experimental tests with the prototype. Some options under test
consist in the use of rings or cylinders (see Fig. \ref{field})
placed at fixed positions around the central stick and at certain
intermediate voltages. Another option being used in combination
with the previous structures is a resistive conic layer placed at
the end of the metallic rod in contact with the small sphere. For
appropriate dimensions of the cone, it provides automatically in
its surface the correct voltage gradient along the first
millimeters away from the sphere, the critical region where the
avalanche occurs. More sophisticated ideas are envisaged for the
future, an example being charging systems like the one used in
electrostatic accelerators, using a series of small metallic balls
on an insulator chain.

\section{Supernova detection and other physics goals}

The detection of neutrino-nucleus interaction opens the way to
other very interesting physics goals. In particular, due to the
coherence of the interaction, modest detection rates can be
achieved by relatively small amount of target material. Neutrinos
coming from a supernova could produce between 600 and 1900 events
in a spherical TPC of 4 m of radius filled with Xe at 10 bar, by
no means a rare event search\cite{Giomataris:2005ver}. A network
of smaller spherical TPCs, distributed around the world has been
proposed \cite{Giomataris:2005ver} as a very efficient supernova
detector, with very simple operation and low cost. It has been
also noted that the neutrino-nucleus interaction may give
information about particular properties of this particle, like the
neutrino charge radius\cite{Bernabeu} and other non-standard
properties of the neutrinos\cite{Barranco}. On the other hand,
more conventional neutrino-electron interaction in the gas can
also be considered\cite{Gounaris:2004ji}, and very low energy
neutrino oscillations could be measured\cite{Giomataris:2003bp}
using a tritium source, which could have sensitivity to
$\theta_{13}$. This type of setup would have a high sensitivity to
the neutrino magnetic moment\cite{McLaughlin:2003yg}. Another
by-product of the spherical TPC is to use it as a neutron detector
by partially (or totally) filling it with He-3. That set-up would
be able to do neutron spectrometry in very low neutron fluxes,
being extremely interesting to characterize neutron backgrounds in
underground laboratories.

\end{document}